\documentclass{elsart}
\usepackage{epsfig}
\usepackage{amssymb}
\usepackage{amsmath}
\usepackage{subfigure}
\usepackage{float}
\usepackage{color}
\usepackage{epsf}

\begin{document}
\begin{frontmatter}

\title{Tracked pellets -- a way to improve the efficiency of charmonium
studies}

\author[ISV]{{\"O}.~Nordhage\corauthref{cor}}\ead{orjan.nordhage@tsl.uu.se},
\author[ISV]{I.~Lehmann\thanksref{label1}},
\thanks[label1]{Current address: Department of Physics \& Astronomy,
University of Glasgow,\\  Glasgow, G12 8QQ, Scotland, UK.}
\author[ISV]{J.~Lith},
\author[TSL]{C.-J.~Frid\'en}, and
\author[ISV]{U.~Wiedner}

\corauth[cor]{Corresponding author.}
\address[ISV]{Department of Nuclear and Particle Physics, Uppsala
University, Box 535, SE-751~21 Uppsala, Sweden}
\address[TSL]{The Svedberg Laboratory, Uppsala University, Box 533,
SE-751~21 Uppsala, Sweden}

\begin{abstract}
We investigate the possibility of tracking individual hydrogen
micro-spheres from an internal pellet target. Such a method aims to
provide the primary vertex of a reaction to within about 100$\,\mu$m,
without utilizing any detector response. Apart from background
considerations the knowledge of the reaction vertex may be essential
for the reconstruction of many physics channels. This is in
particular true for the study of the $\mathrm{\Psi}(3770)$ decay
into D-mesons planned at the PANDA detector at the future FAIR
facility. Here the reconstruction of displaced vertices is
especially difficult since neutral particles are involved.

Studies with a pellet target at The Svedberg Laboratory, Uppsala,
show the technical feasibility of a tracking system utilizing fast
CCD line-scan cameras. Simulations for the reaction
$\bar{\mathrm{p}}\mathrm{p}\rightarrow\mathrm{\Psi}(3770)
\rightarrow\mathrm{D}^{+}\mathrm{D}^{-}$ prove the large impact such
a system would have on the data taking and reconstruction at PANDA.
\end{abstract}


\end{frontmatter}

%\tableofcontents

%%%%%%%%%%%%%%%%%%%%%%%%%%%%%%%%%%%%%%%%%%%%%%%%%%%%%%%%%%%%%%%%%%%
\section{Introduction}
\label{intro} Charmonium physics has been shown to be a powerful
tool in the understanding of the strong interaction and hadronic
structure. Probes like antiprotons are unprecedented due to the
gluon-rich environment, which they assure. In the PANDA experiment
\cite{TPR} at the future FAIR facility antiprotons in the momentum
range between $1.5\,$GeV/$c$ and $15\,$GeV/$c$ will interact with an
internal target. Most of the measurements require a proton target
and the use frozen micro-spheres, ``pellets'', of hydrogen with a
typical size of $30\,\mu$m is planned~\cite{Pellet,ON05}. To create
pellets hydrogen liquid is forced through a nozzle and, under the
influence of vibrations, the liquid jet breaks up into droplets.
These are then injected into vacuum during which freezing occurs and
the result is a stream of pellets (up to $10^{4}\,$pellets/sec)
reaching the interaction point.

Such pellets have the advantage of being discrete objects separated
by a few millimeters distance, and hence individually traceable.
This, however, poses technical challenges which, in a first approach
using a light-guide system and preamplifiers, were difficult to
overcome~\cite{LG}. We propose the use of fast CCD line-scan cameras
which we believe is sufficiently developed to be of use. As typical
experimental set-ups such as PANDA leave almost no space close to the
interaction point, arrays of cameras and lasers must be arranged
above and below the experiment. The position and time that the
individually-traced pellet passes the beam has then to be
extrapolated from information gathered up to 2\,m away.

In the first part of this document we show how the small amount of
light reflected from a pellet can be registered in a commercial
100\,kHz CCD line-scan camera. In the second part we investigate how
such tracked pellets would improve the fraction of D-meson decays
that can be observed as displaced vertices in PANDA.

%%%%%%%%%%%%%%%%%%%%%%%%%%%%%%%%%%%%%%%%%%%%%%%%%%%%%%%%%%%%%%%%%
\section{Pellet Tracking}
\label{sec:TM}

\begin{figure}[thb]
   \begin{center}
     \includegraphics[width=0.8\columnwidth]{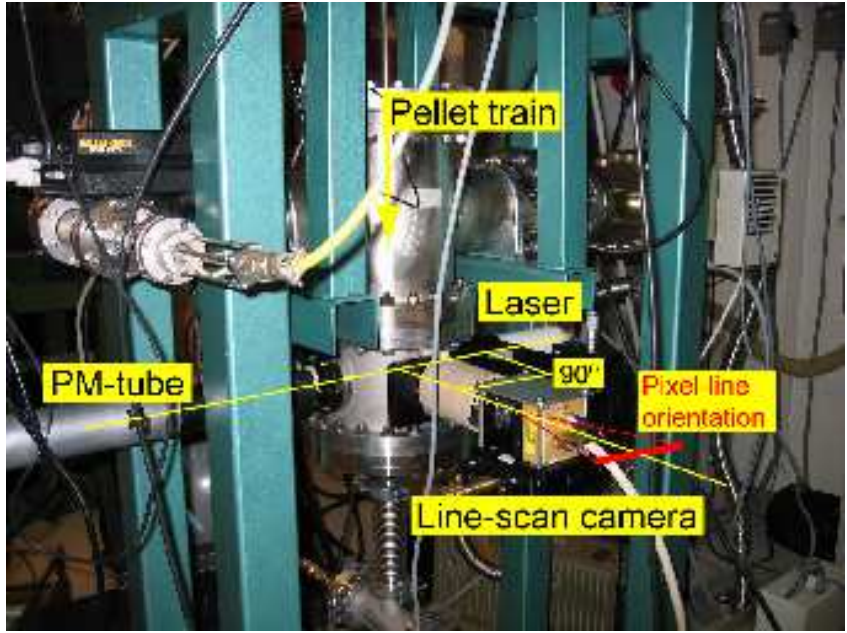}
     \caption{Pellet-Test Station equipped with a line-scan camera
       perpendicular to the laser beam (horizontal right to left) and
       pellet stream (vertical). The photo-multiplier tube is
       equipped with a collimator and optics and serves as an
       independent rate monitor. The system is installed 2.3\,m below
       the place of pellet production (the vacuum injection capillary).}
   \label{fig:LSC_setup}
  \end{center}
\end{figure}

The novel idea is to realize a tracking system using reflected laser
light from the pellets seen by fast CCD line-scan cameras located
above and below the interaction point. Four cameras should be
positioned in pairs, two above and two below the interaction point,
with 90$^{\circ}$ relative angle to each other. This way two parallel
planes are defined, each providing the pellet-position in $x$ and
$z$. The pellet path is extracted from the position information from
each plane and relative timing. A feature of the current pellet stream
is a rather big difference in velocities of the pellets. Thus {\it a
priori} it cannot be excluded that pellets may overtake each other
between the two observation planes.  The consequent problem of
assigning a unique number to each pellet can be solved by having an
additional camera at almost any place in the pellets' path. Also the
system will naturally provide rates and distributions for the
statistical analysis.

A prototype was built at the Pellet-Test Station (PTS) at
The Svedberg Laboratory, Uppsala, Sweden. The PTS is an independent
pellet target system which is, to a large extent, a direct copy of the
WASA pellet target system~\cite{Pellet,WASA}. The pellet generator
located at the top of the system is apart from minor improvements
identical. The lower vacuum system is designed to perform vacuum and
distribution studies in a configuration resembling a typical internal
experiment. The prototype of the tracking system was attached to
windows of the lower vacuum system and consists of an AViiVA line-scan
camera with 512 pixels (each $14\times14\,\mu$m$^{2}$) and a laser
placed perpendicularly to it (see Figure~\ref{fig:LSC_setup}). A laser
of 35\,mW with a width of 5\,mm and height of about $100\,\mu$m was
used, covering the whole width of the distribution and reducing the
probability of having two pellets simultaneously inside the laser
light down to the percent-level. The line read-out speed and dead time
of the camera have been determined experimentally to by
$90,909\pm1$\,Hz and $3.37\pm0.02$\,\%, respectively~\cite{Lith}.

\begin{figure}[thb]
  \begin{center}
\subfigure{\includegraphics[width=0.495\columnwidth]{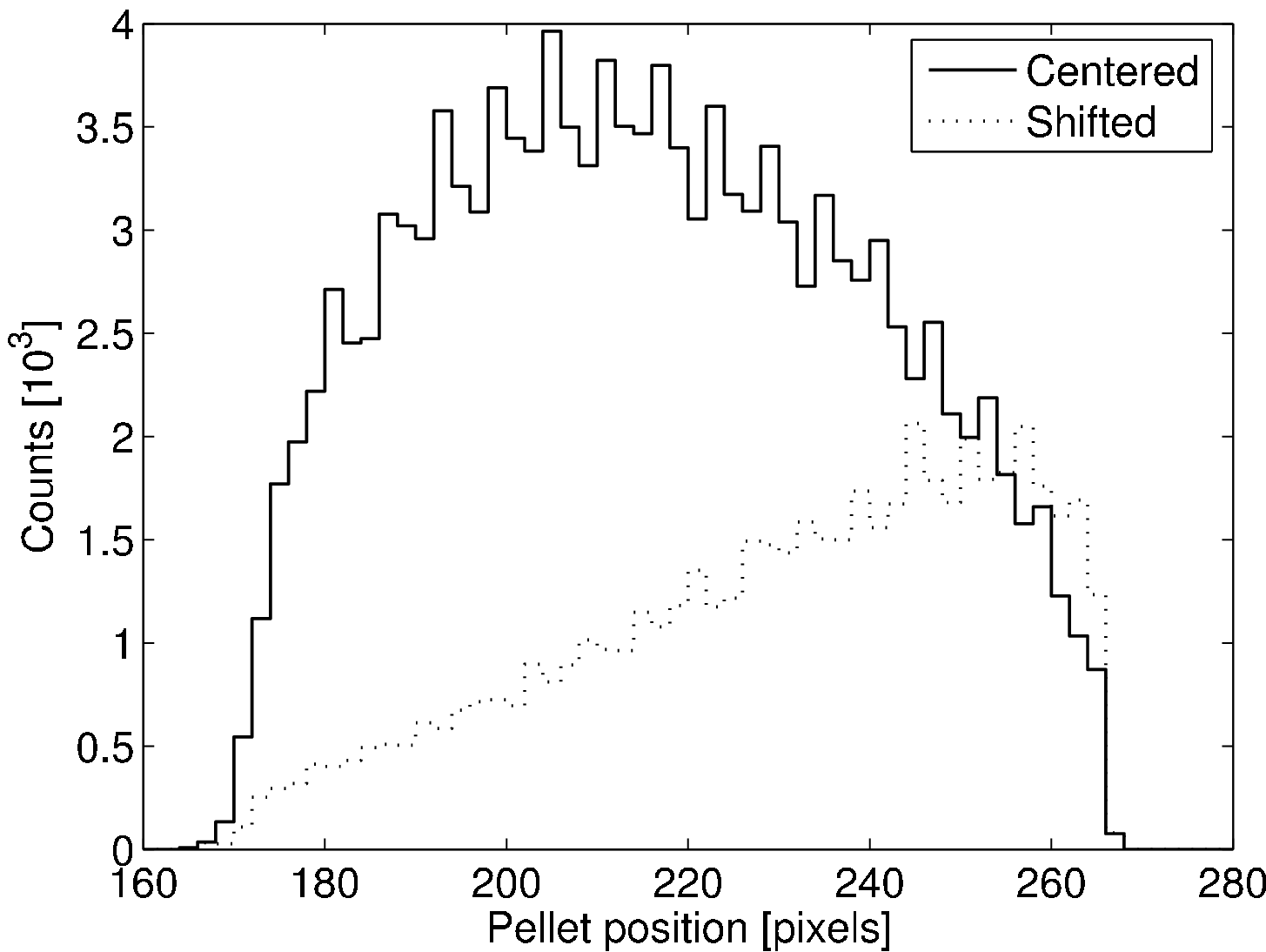}}
\subfigure{\includegraphics[width=0.495\columnwidth]{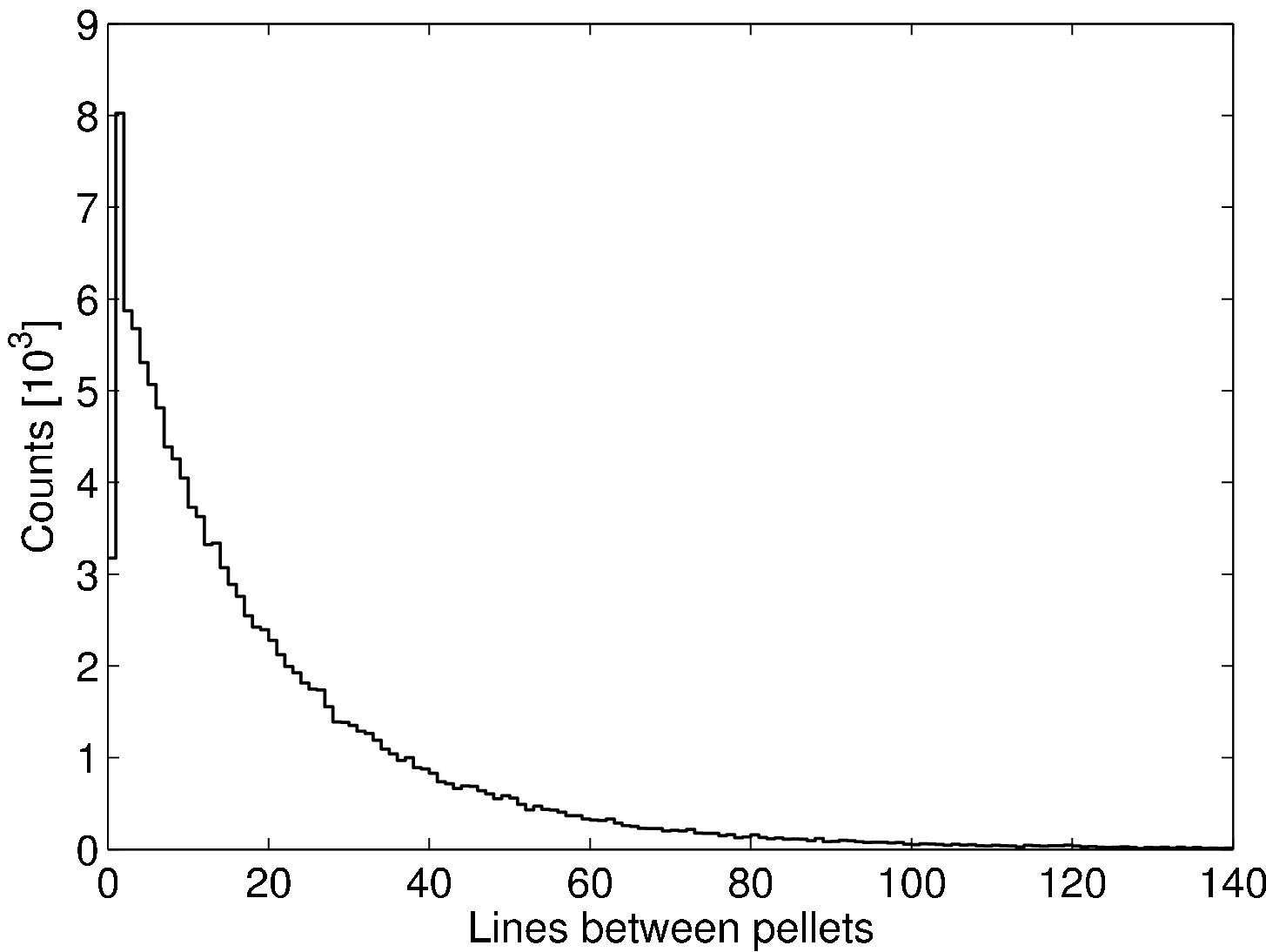}}
  \caption{Left: Spatial distribution of the pellets for two different
    run conditions where the total widths of 102 pixels correspond to
    3.8\,mm. Right: Number of lines between two subsequent pellets for
    the centered pellet stream. Each line corresponds to $11\,\mu$s.}
   \label{fig:PelletData}
  \end{center}
\end{figure}

The test measurements showed the feasibility of detecting individual
pellets and determining their position with such a system. The light
yield under 90$^\circ$ was sufficient to be detected in the pixels
of the camera. Standard optics (focal length 50\,mm and f=1.4) at a
focal distance of 184\,mm and magnification of 0.38 were used. The
distribution of pellets 2.3\,m below their place of production is
shown in Figure~\ref{fig:PelletData} (left) for two different cases.
A circular skimmer of 2.0\,mm diameter is situated 1.07\,m above the
measurement plane and cuts the tails of the distribution. A centered
pellet stream (maximized count rate) produces a symmetric
distribution (solid line). A pellet stream moved to the right
results in smaller count rate and an asymmetric pellet distribution
(dotted line), while the boundaries (given by the fixed skimmer)
remain constant. One pixel corresponds to 37$\,\mu$m, thus the total
pellet-stream width is calculated to be 3.8\,mm which is in good
agreement with the expected width of 3.7\,mm from geometrical
considerations.

The right side of Figure~\ref{fig:PelletData} shows the time
distribution between two subsequent pellets. Each line corresponds to
$11.0\,\mu$s, during which the light in the pixels is sampled. Two
pellets are seen within a single line in 2.5\% of the scans taken.
But in most cases this poses no problem for the tracking, as the
pellets can be distinguished due to separated positions. 6.2\% of the
pellets appear in the subsequent line, while 50\% are detected within
less than $180\,\mu$s. These results confirm that the distribution of
pellets is governed by the statistical selection of pellets through
the skimmer and the spread in velocities, rather than by the frequency
which is used to generate the original droplets (about 80\,kHz).

The light intensity profile recorded with the camera indicates that
we could be at the limit of the sensor's sensitivity. The fact that
the independently measured rates in a carefully adjusted counter
coincide, however, make a significant inefficiency unlikely. In our
case more light reflected into the camera would have helped to study
the effect. Another advantage of having superfluous light would be
that one could close the lens aperture to avoid smearing due to a
limited range of focus. A stronger laser arranged to angles closer
to 180$^\circ$ with respect to the camera (i.e. shining almost into
the camera) may help in that respect, as observations indicate a
large transparency of those pellets.

All bulky target equipment must be placed outside the
detector at PANDA. Therefore, the distance between the two above-mentioned
camera planes will be $\gtrsim4\,$m and the time delay for a
particular pellet position will be $\sim0.1\,$s due to the pellet
travel time. A latency of this magnitude causes the information to
enter at a higher level in the trigger scheme, but can be
efficiently used for any offline analysis.

The position resolution of the reconstructed vertex will depend on
several factors. The pixel size of $14\,\mu$m leads to a resolution
of $\sigma_\mathrm{pix}=37\,\mu$m in the plane of the pellets. The
largest uncertainty will be introduced by the alignment of the
cameras. This depends upon the construction and the calibration
procedures. It is assumed that such an absolute alignment could be
done within $\sigma_\mathrm{align}=50\,\mu$m.  Both uncertainties
are independent and are added quadratically. For our approach we
assume a symmetric arrangement with respect to the beam line. Thus
we calculate the $x$-coordinate at the interaction point simply as
the mean of the values for each plane (denoted by 1 and 2)
$x_\mathrm{ip}=(x_{1}+x_{2})/2$. The same applies for the
$z$-coordinate.  As the uncertainties are equal in both planes we
obtain the overall uncertainty of the pellet-position at the
interaction point for the $x$ and $z$-coordinate
\begin{equation}\nonumber
\sigma^\mathrm{pellet}_{x,\,\mathrm{ip}}=
\sigma^\mathrm{pellet}_{z,\,\mathrm{ip}}=
\sqrt{\big(\frac{1}{2}\big)^{2} \times
2\times(\sigma_\mathrm{pix}^2+\sigma_\mathrm{align}^2)}=44\,\mu\mbox{m}.
\end{equation}
The third coordinate along the direction of travel of the pellets
(decreasing $y$) is determined by the time resolution, which is
equal to the active time of the camera. The maximum error we would
expect here is 11$\,\mu$s which corresponds to well below a
millimeter of pellet travel since the typical speed is
$\gtrsim50\,$m/s.

%%%%%%%%%%%%%%%%%%%%%%%%%%%%%%%%%%%%%%%%%%%%%%%%%%%%%%%%%%%%%%%%%
\section{Displaced Vertices}

One of the physics goals of PANDA is charmonium spectroscopy.
Charmonium spectroscopy above the
$\mathrm{D}\bar{\mathrm{D}}$-threshold has hardly been investigated
because D-mesons are hard to select in an experiment due to their
short lifetime. In PANDA we have several measures to identify D-mesons
and one tool is to directly recognize the displaced vertex caused by
the decay. This assumes that the charged particles reach the innermost
tracking detectors. At PANDA this is a micro-vertex detector (MVD)
arranged in a barrel geometry of layers with silicon pixels and
silicon strips. Any charged particle will be detected here and the
resolution for the decay $\mathrm{D}^{\pm}\rightarrow
\mathrm{K}^{\mp}\pi^{\pm}\pi^{\pm}$ is estimated to be about
$50\,\mu$m~\cite{AS}. Such information could enter early into the
trigger scheme and allow the pre-selection of potential ``hidden-charm
candidates''.

A much more powerful tool to identify D-meson decays is the
reconstruction of their decay length, through knowledge of the primary
reaction vertex. Often this primary vertex cannot be reconstructed
using the reaction products, as they tend to either be too short-lived
or neutral particles. Tracking individual pellets, as proposed here,
yields a completely independent determination of this primary
vertex. In conjunction with the charged particle tracking, such a
system would be used both for the physical analysis and for background
rejection.

We investigated the impact of such an approach by studying the
benchmark process
\begin{equation*}
\bar{\mathrm{p}}\mathrm{p}\rightarrow\mathrm{\Psi}(3770)\rightarrow
\mathrm{D}^{+}\mathrm{D}^{-},
\end{equation*}
where the hidden-charmed $\mathrm{\Psi}$ is produced at rest in the
center-of-mass system, corresponding to an antiproton momentum of
6.59\,GeV/$c$. In order to be able to quantify the impact of the
pellet-tracking method we compare two scenarios, one with pellet
tracking and one without.

Decays of charged particles are often accompanied by neutral particles
in the final state (e.g.\ $\mathrm{D}^{+}\rightarrow
\bar{\mathrm{K}}^{0}\pi^{+},\, \bar{\mathrm{K}}^{ 0}\pi^{+}\pi^{0},$
or $\bar{\mathrm{K}}^{0}\pi^{+}\pi^{+}\pi^{-}$). Therefore the
reconstruction of the vertex is not possible from the observed charged
decay products. To build an efficient trigger on such decays one needs
to identify a significant kink in the particle track.  However, one
problem in identifying kinks from the D-mesons in the reaction above
is the lifetime $\tau=1040\,$fs \cite{PDG} leading to an average
travel length of only $\simeq0.5\,$mm. In addition, demanding an
exclusive detection of the D-decay particles would be rather a
inefficient constraint since the decay modes that contain at least one
neutral particle add up to a substantial fraction. Already foreseen is
to reconstruct vertices from the MVD response to be used in a
first-level trigger, to see whether two particles have decayed within
some narrow time window and if their decay points are separated
\cite{WK}. In such a case the knowledge of the primary vertex would be
of great help and therefore the discrete nature of pellets is an
important advantage when compared to a typical cluster-jet target with a
continuous target density and a much larger spatial extension. We will
not look further on a comparison between cluster targets and a pellet
target, but rather compare tracked pellets to untracked pellets.

To explore this, we introduce the convolution between the antiproton
beam and target as the ``possible volume of primary interactions'',
V$_\mathrm{prim}$. For the antiproton beam at the target location we
can assume that the combined effects of (stochastic) cooling and
heating due to target and intra-beam scattering will result in a
Gaussian distribution with rms widths $\sigma_{x}$ and $\sigma_{y}$
in the horizontal and vertical direction, respectively. A single
pellet can be assumed to be spherical and of diameter $30\,\mu$m. If
we randomly generate primary vertices inside the pellet as many as
98\% of the D-mesons will decay outside.

\begin{figure}[thb]
  \begin{center}
  \subfigure[Primary vertices (untracked
pellets).]{\includegraphics[width=0.49\columnwidth]{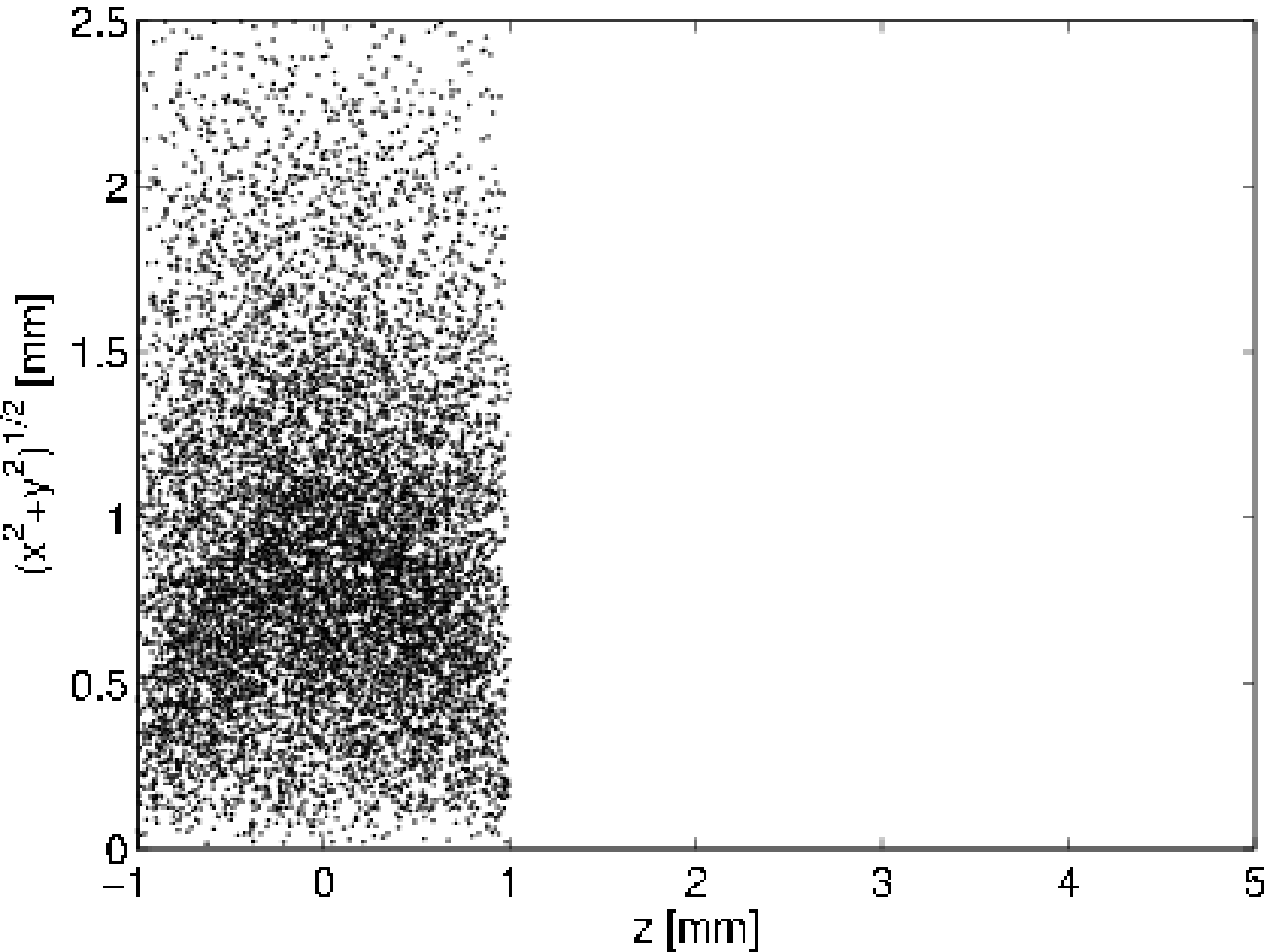}}
  \subfigure[Primary vertices (tracked
pellets).]{\includegraphics[width=0.49\columnwidth]{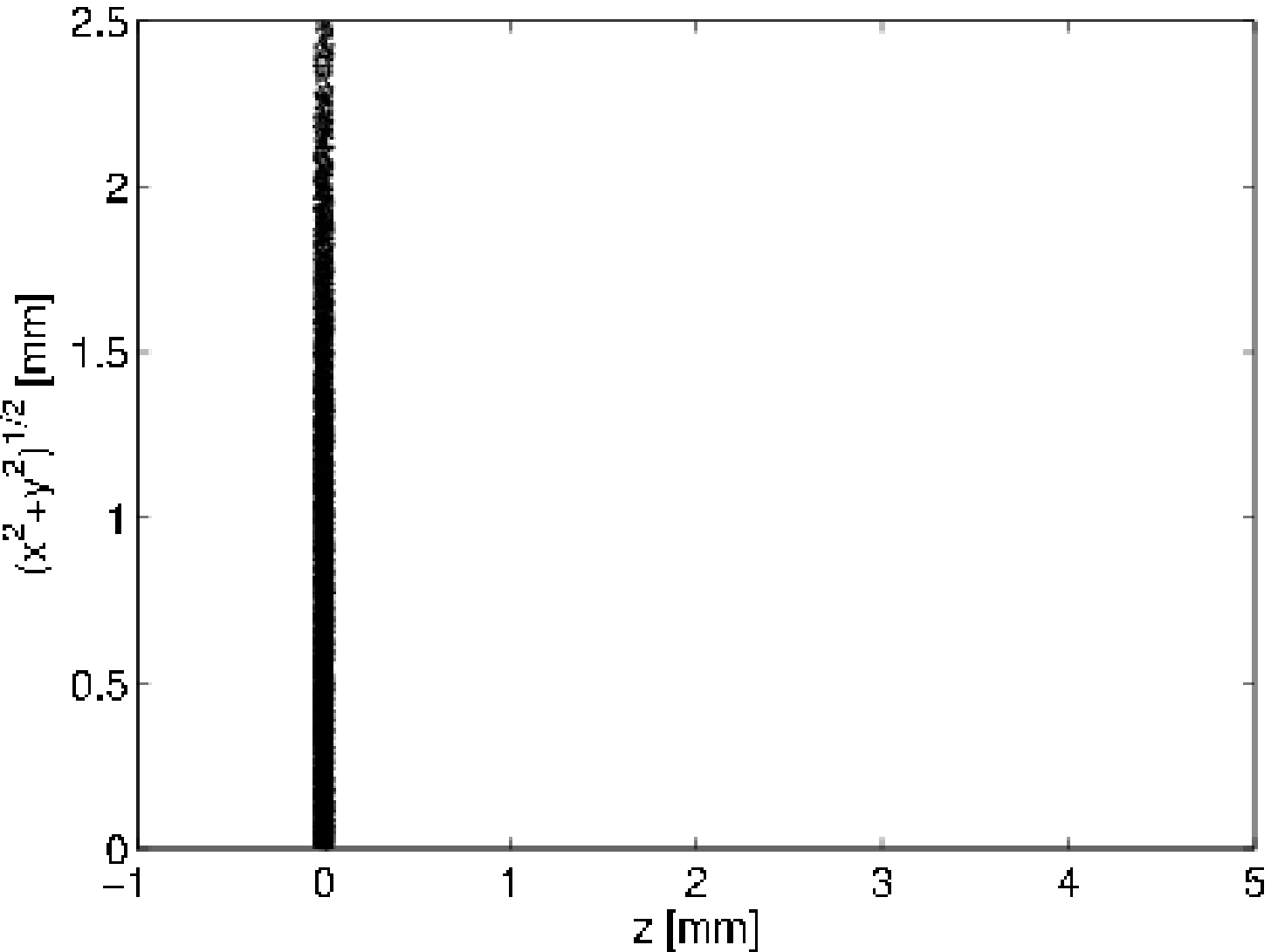}}
\\
  \subfigure[D-decay vertices (untracked
pellets).]{\includegraphics[width=0.49\columnwidth]{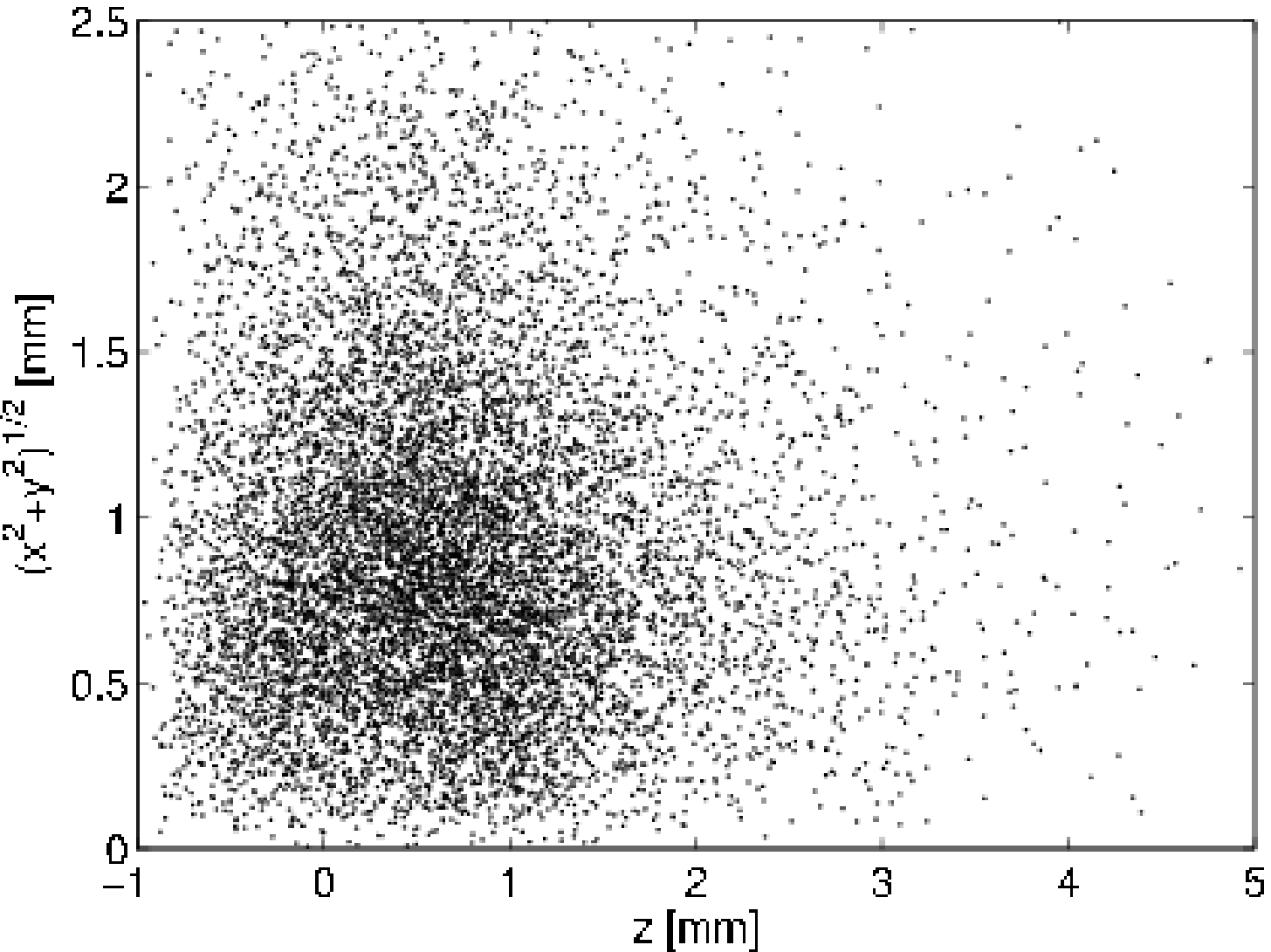}}
  \subfigure[D-decay vertices (tracked
pellets).]{\includegraphics[width=0.49\columnwidth]{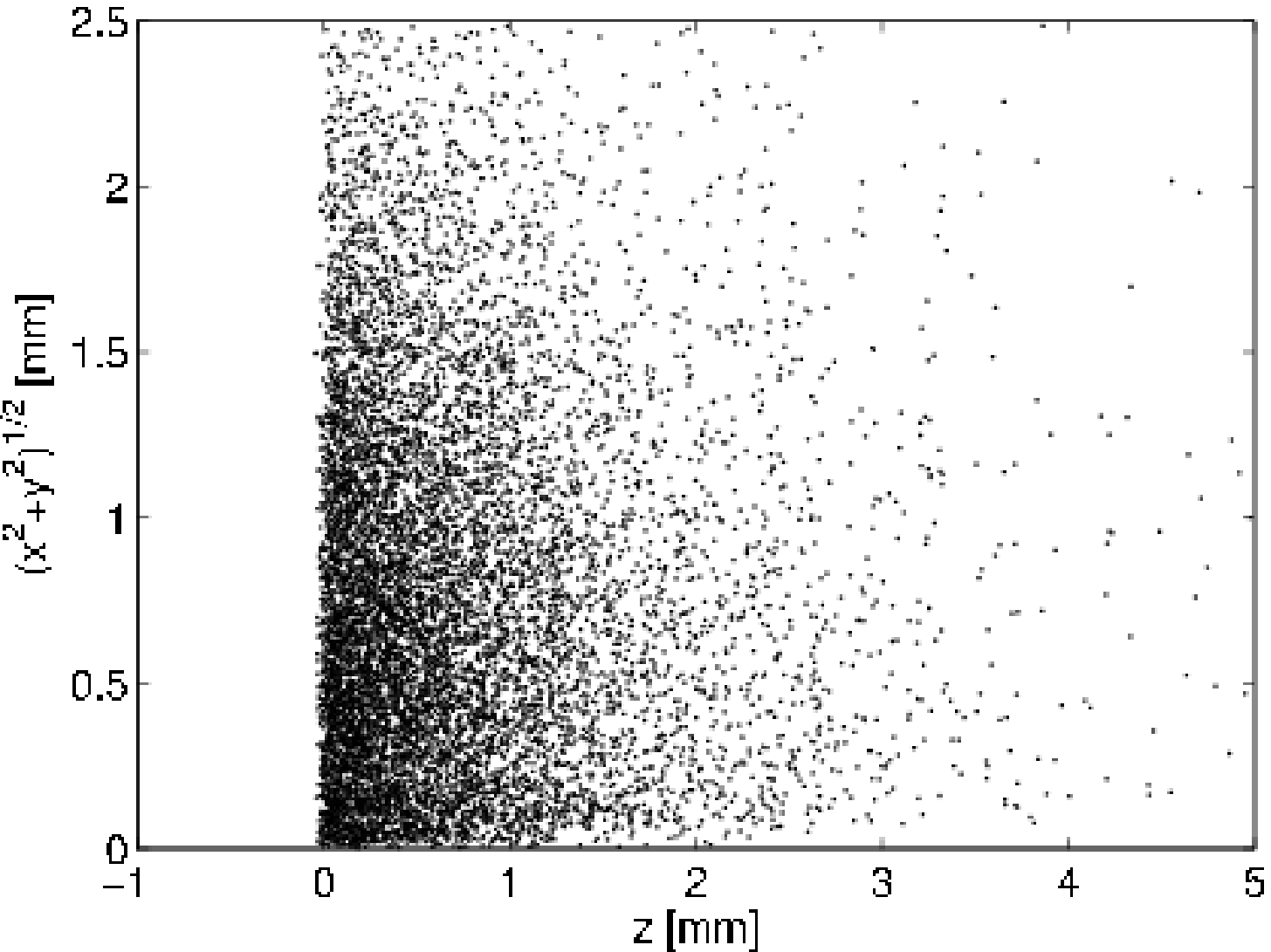}
}
   \caption{In a) and b) 10,000 randomly generated primary vertices
   are shown for untracked and tracked pellets, respectively. In c)
   and d) the corresponding distributions of secondary vertices from
   the benchmark channel are plotted. The beam comes in from -z and
   meets the center of the target at the origin. The vertical axis
   shows the transverse distance to that point.}
   \label{fig:Vertices}
  \end{center}
\end{figure}

We will consider primary vertices V$_\mathrm{prim}$ based on an
antiproton beam with $\sigma_{x}=\sigma_{y}=1\,$mm and:
\begin{itemize}
     \item a randomly spread distribution in the $xz$-plane within a
     circle\footnote{The average density distribution is in fact almost
     a constant within the circular area defined by the skimmer. This
     can also be seen from the projection shown in
     Figure~\ref{fig:PelletData} (left). For further details see
     Ref.~\cite{ON_STORI}.} of diameter 2\,mm, which is to correspond
     to untracked pellets, and
     \item a trace-tracked pellet for which the $xz$-position is
     determined to within a diameter of 0.1\,mm.
\end{itemize}
For convenience, we take the center of the region of primary
interactions as reference frame, i.e.\ (0,0,0), such that a direct
comparison between the above-mentioned cases is possible. In
Figure~\ref{fig:Vertices} we see 10,000 randomly generated
vertices. Figures a) and b) show the primary
$\bar{\mathrm{p}}\mathrm{p}$-vertices (corresponding also to the
$\mathrm{\Psi}$-decay points) whereas c) and d) show the distributions
of the D-meson decay points. We observe that the separation of primary
and secondary vertices will be much more efficient with pellet
tracking, but still possible without.

To detect neutral D$^{0}$-mesons which decay even faster than
D$^{\pm}$ it will be of even greater importance to know the primary
vertex. Otherwise it will be difficult -- if not impossible -- to
distinguish prompt reactions like $\bar{\mathrm{p}}\mathrm{p}
\rightarrow \pi^{+}\mathrm{K}^{-} \mathrm{K}^{0}_\mathrm{S}$ from
those where the resonant decay $\mathrm{D}^{0}\rightarrow
\pi^{+}\mathrm{K}^{-}$ is involved. It should also be pointed out that
the secondary vertices will be shifted in the longitudinal direction
as the beam momentum and therefore the Lorentz boost increases. This
will, of course, facilitate the vertex separation.

We define the longitudinal displacement of the secondary and primary
vertex $ d \equiv z_\mathrm{D} -
z_{\mathrm{V}_\mathrm{prim}}^\mathrm{max}.$ Here $z_\mathrm{D}$ and
$z_{\mathrm{V}_\mathrm{prim}}^\mathrm{max}$ are the $z$-coordinates of
the D-meson decay point and maximum value of the possible primary
vertex, respectively. The uncertainty of $d$ will depend on the
uncertainty of $z_\mathrm{D}$, thus the MVD resolution, and the
uncertainty of $z_{\mathrm{V} _\mathrm{prim}}^\mathrm{max}$, which we
assume to be 10\%. In Table~\ref{t:disp_res} the percentage of
D-mesons that one can identify by their decay (i.e.\ having a
displacement $d>0$) is shown. We find that the identification can be
improved by a factor of 4 to 5 using tracked pellets in comparison to
the conventional approach.

\begin{table}[htb]
\begin{center}
\begin{tabular}{c|cc}
\hline \hline
Confidence & \multicolumn{2}{|c}{D-mesons with $d>0$ $[$\%$]$} \\
level & $z_{\mathrm{V}_\mathrm{prim}}^\mathrm{max}=1\,$mm &
$z_{\mathrm{V}_\mathrm{prim}}^\mathrm{max}=50\,\mu$m \\
& (Untracked) & (Tracked) \\
  \hline
1$\sigma$ & 20 & 83  \\
2$\sigma$ & 16 & 76  \\
3$\sigma$ & 13 & 69  \\
\hline \hline
\end{tabular}
\end{center}
\caption{Fraction of D-mesons with a positive (thus distinguishable)
displacement.} \label{t:disp_res}
\end{table}

%%%%%%%%%%%%%%%%%%%%%%%%%%%%%%%%%%%%%%%%%%%%%%%%%%%%%%%%%%%%%%%%%
\section{Conclusions}

We have shown the feasibility of tracking individual pellets using
laser light reflected into a fast line-scan camera. In a test
experiment the rates from a conventional system and the expected
transverse distribution of pellets could be well reproduced. By
using several cameras and lasers we propose a system which could
provide a three-dimensional vertex point at any experiment using
such a target. If used at the future PANDA experiment we expect to
obtain a resolution at the interaction point of the order
$\lesssim100\,\mu$m in the two perpendicular directions and $\le
1\,$mm in the parallel direction, with respect to the pellet stream.

A pellet-tracking system would in particular be advantageous in the
study of events with hidden charm, where the reconstruction of the
D-meson decay length is a prerequisite. It would also be used online
to reduce the data stored in a late stage of the data-acquisition
system and for the rejection of background.

We have demonstrated on the benchmark reaction
$\bar{\mathrm{p}}\mathrm{p}\rightarrow\mathrm{\Psi}(3770)
\rightarrow\mathrm{D}^{+}\mathrm{D}^{-}$ that the efficiency of the
reconstruction of D-meson decays can be increased by a factor of
$\sim4$ using a system that tracks pellets. This shows that such an
approach will substantially improve the data quality.

\section*{Acknowledgements}
We would like to thank all the staff of The Svedberg Laboratory for
their help with all kind of odds and ends, especially Kjell
Fransson, Gunnar Norman, and the workshop personnel. We thank
M.~Murray and C.~Shearer for revising the document. The authors
greatly acknowledge the G{\"o}ran Gustafsson Foundation for granting
the application on pellet tracking. One of us ({\"O}.N.) also wishes
to thank GSI, Darmstadt, Germany, for the financial support.


\begin{thebibliography}{99}


\bibitem{TPR} PANDA Collaboration, \emph{Technical Progress Report}
(2005).

\bibitem{Pellet}
   B.\ Trostell,
   %Vacuum injection of hydrogen micro-sphere beams,
   Nucl.\ Instr.\ and Meth.\ in Phys.\ Res.\ A\ \textbf{362}
   (1995) 41; C.~Ekstr{\"o}m \emph{et al.}, Nucl.\ Instr.\ and Meth.\ A\
\textbf{371} (1996) 572.

%\bibitem{PT1} C.~Ekstr{\"o}m \emph{et al.},
%\emph{Hydrogen pellet targets for circulating particle beams}

\bibitem{ON05} {\"O}.~Nordhage \emph{et al.},
Nucl.\ Instr.\ and Meth.\ A\ \textbf{546} (2005) 391.

\bibitem{LG} S.~Goldberg and L.~Gustafsson, TSL/ISV-Report 95-0122,
   Uppsala University (1995); P.~Wasylczyk, Internal note TSL,
   Uppsala University (1996).

\bibitem{WASA} C.~Ekstr{\"o}m \emph{et al.},
   Phys.\ Scripta T99 (2002) 169;
   J.~Zabierowski \emph{et al.}, Phys.\ Scripta T99 (2002) 159.

\bibitem{Lith} J.~Lith, Master's Thesis, Uppsala University (2006);
M.~Schult, Project Work, Uppsala University (2005).

\bibitem{AS} A.~Sokolov, Ph.D.~Thesis, Universit\"{a}t
Giessen (2005).

\bibitem{PDG} S.~Eidelman \emph{et al.} (Particle Data Group),
Phys.~Lett.~B
\textbf{592}, 1 (2004) and 2005 partial update for edition 2006
(URL: http://pdg.lbl.gov).

\bibitem{WK} W.~K{\"u}hn, Personal communication (2006).

\bibitem{ON_STORI} {\"O}.~Nordhage \emph{et al.}, Conference
Proceedings STORI'05, J{\"u}lich, Vol 30, (2005) 389.

\end{thebibliography}
\end{document}